\documentclass[12pt,titlepage]{article}
\usepackage{fullpage}
\usepackage{lipsum}
\usepackage[T1]{fontenc}
\usepackage[english]{babel}
\usepackage[utf8x]{inputenc}
\usepackage{graphicx}
\usepackage{amsmath}
\usepackage{amsfonts}
\usepackage{hyperref}
\usepackage{bbding}
\providecommand{\keywords}[1]{\textbf{\textit{Keywords---}} #1}
\title{Finding of k in Fagin's R. Theorem 24 }
\author{Algirdas Antano Maknickas
\\
\small Department of Information Technologies, \\[-0.8ex]
\small Vilnius Gediminas Technical University, \\[-0.8ex]
\small         Sauletekio al. 11, Vilnius, Lithuania \\[-0.8ex]
\small               \texttt{algirdas.maknickas@vgtu.lt}  
}
\begin{document}
\maketitle 
\begin{abstract}
By using of analytical multi-logic expresses in conjunction with non-deterministic Turing machine the proposition was proved that algorithm of deterministic Turing counter machine of polynomial time complexity can be decreased to the algorithm of linear time complexity in non-deterministic Turing counter machine.  Furthermore, it was shown that existence of reduction of polynomial time complexity to the linear time complexity by switching  from deterministic to non-deterministic Turing machine for string recognition imply P equals to NP. Thereto, analytical generation functions of higher order logic were used for finding of k value in Fagin's R. Theorem 24. 
\end{abstract}
\section*{Introduction}

\textit{Importance.} Computational complexity theory plays an important role in modern cryptography \cite{Goldreich2000}.  The security of the Internet, including most financial transactions, depends on complexity-theoretic assumptions such as the difficulty of integer factoring or of breaking DES (the Data Encryption Standard).  If P=NP, these assumptions are all false.  Specifically, an algorithm solving 3-SAT in $n^2$ steps could be used to factor 200-digit numbers in a few minutes. The much more detail problem description and importance can be found in \cite{Cook}.

\textit{Turing machines.} The standard computer model
in computability theory is the Turing machine, introduced by Alan Turing in 1936
\cite{Turing1936}. A Turing machine $M$ consists of a finite state control (i.e., a finite program) attached to a read/write head moving on an infinite tape. The tape is divided into squares, each capable of storing one symbol from a finite alphabet $\Gamma$ that includes the blank symbol $b$. Each machine $M$ has a specified input alphabet $\Sigma$, which is a subset of $\Gamma$, not including the blank symbol $b$. At each step in a computation, $M$ is in some state $q$ in a specified finite set $Q$ of possible states. Initially, a finite input string over $\Sigma$ is written on adjacent squares of the tape, all other squares are blank (contain $b$), the head scans the left-most symbol of the input string, and $M$ is in the initial state $q_0$. At each step $M$ is in some state $q$ and the head is scanning a tape square containing some tape symbol $s$, and the action performed depends on the pair (q, s) and is specified by the machine’s transition function (or program) $\delta$. The action of deterministic Turing machines consists of printing a symbol on the scanned square, moving the head left or right one square, and assuming a new state \cite{Cook}. Obviously, a deterministic Turing machines are related to binary logic and in most cases strings researchers offered to represent this strings in binaries. 

\textit{Turing machines over $\mathbb{C}$.} A Turing machine over $\mathbb{C}$ first was proposed by Blum, Shub, and Smale \cite{Blum1989} and the associated theory BCSS (or Blum-Cucker-Shub-Smale) is exposed in \cite{Blum1997}. As mentioned in \cite{Smale1998}, a Turing machine over $\mathbb{C}$ has as inputs a  finite string $( ..., x_{−1}; x_0; x_1, ... )$ of complex numbers and the same for states and outputs.  Computations on states include arithmetic operations and shifts on the string. The  size  of an input is the  number  of elements  in the input  string.   The  time  of a computation is the number of machine operations used in the passage from input to output. Thus a polynomial time algorithm over $\mathbb{C}$ is well-defined. Note that all that has been said about the machines  use only the structure of $\mathbb{C}$ as a  field and hence the machines  make sense over any  field. In particular if the  field is $\mathbb{Z}_2$ of two elements, we have the deterministic Turing machines.

The action of non-deterministic Turing machines consists of printing a symbol on the scanned square, moving the head left (1, 2, 3, ...) or right (1, 2, 3, ...) square, and assuming a new state. So, a non-deterministic Turing machines are related to multivalued logic and for string coding integer numbers can be used.  Furthermore, if we can formulate complex Turing machines as non-deterministic Turing machines manipulating on the  field $\mathbb{Z}_n$ of n elements \cite{Maknickas2015}, so we can do it on the field of complex numbers with fixed length  $z_n\in\mathbb{C}, |z_n|=1,n\in\mathbb{Z}^+$.

\textit{Complex binary and multivalued logic.} The description of complex binary logic is described in section "Lemmas of binary logic". The  complex multivalued logic is the extension of complex binary logic and is described in section "Lemmas of multivalued logic". 

\textit{Aims and scopes.} The aim of this paper is the proof of Fagin's Theorem 24 \cite{bib:Fagin} which claims:

TEOREM. \textit{The following two statements are equivalent:}
\begin{enumerate}
\item $NP = P$.
\item \textit{There exists a constant $k$ such that, for every countable function $T$
with $T\left(l\right) \geq l + 1$ for each l and for every language $A$ which is recognized
by a non-deterministic one-tape Turing machine in time $T$, the language $A$ is
recognized by a deterministic one-tape Turing machine in time $T^k\geq l^k+1$}.
\end{enumerate}
The general proposition is to use in the prof of this theorem complex Turing machine over the field of complex numbers with fixed length  $z_n\in\mathbb{C}, |z_n|=1,n\in\mathbb{Z}^+$ by extending operation with complex numbers from three operation (summation, multiplication and negotiation or $+,-,\times$) to the all $n^n$ operations prescribed over single complex number in complex multivalued logic. 

\section{Notions from automata theory}
Denote the set of complex  numbers $\left\lbrace \exp{(2 \pi/n)}, \exp{(4\pi /n)}, ..., \exp{(2i\pi /n)}\right\rbrace$ 
by $\mathbb{C}$ where $i,n\in\mathbb{Z}^+$ and $i\leq n$, and the set $\left\lbrace \exp{(2\pi\cdot 0 /n)}, ... , \exp{(2(n - 1)\pi /n)}\right\rbrace$  by $z_n$.  If $A$ is a set, then card $A$ is the cardinality of the set. Denote
the set of $k$-tuples $< a_1, ... , a_k >$ of members of $A$ by $A^k$.

When $A$ is a finite set of symbols, then $A^*$ is the set of strings, that is, the finite concatenations $a_1 \diamond a_2 ... \diamond a_n$ of members of A. The length of $a = a_1 \diamond a_2 ... \diamond a_n$ is $n$ (written
$len(a) = n$). If $k \in \mathbb{Z}^+$, then $len(k)$ is the length of the binary representation
of $k$ in complex plane; this corresponds to a convention that we will same time represent positive
integers in binary notation in complex plane. If a set $S \subseteq A^*$ for some finite set $A$, then $S$ is
a \textit{language}.

An \textit{m-tape non-deterministic Turing machine M} is an 8-tuple $<K$, $\Gamma$, $B$, $\Sigma$, $\delta$, $q_0$,
$q_A$, $q_R>$, where $K$ is a finite set (the states of $M$); $\Gamma$ is a finite set (the tape symbols
of $M$); $B$ is a member of $S$ (the blank); $\Sigma$ is a subset of ($\Gamma$ - $\left\lbrace B \right\rbrace $) (the input symbols
of $M$); $q_0$, $q_A$, and $q_R$ are members of $K$ (the initial state. accepting final state, and
rejecting final state of $M$, respectively); and ${\delta}$ is a mapping from ${\left( K - \left\lbrace q_A, q_R \right\rbrace \right)} \times {\Gamma^m}$
to the set of non empty subsets of ${K \times \left( \Gamma - \left\lbrace B \right\rbrace \right)^m} \times {\left\lbrace L, R \right\rbrace ^m}$ (the table of transitions, moves, or steps of $M$). 

If the range of $\delta$ consists of singletons sets, that is, sets with exactly one
member, then $M$ is an \textit{m-tape deterministic Turing machine}. 

An \textit{instantaneous description} of $M$ is a $\left( 2 m + 1\right)$tuple $I =$ $< q;\alpha^1, ... ,\alpha^m;i_1,$ $ ... , i_m >$, where $q \in K$, where $\alpha_j \in \left(\Gamma - \left\lbrace B \right\rbrace \right)^*$, and
where $1 \leq i_j \leq len\left(\alpha^j\right) + 1$, for $1 \leq j \leq m$. We say that $M$ is in state $q$, that
$\alpha^j$ is the non blank portion of the $j$-th tape, and that the $j$-th tape head is scanning
$\left(\alpha^j\right)_{i_j}$, the $j$-th symbol of the string $\alpha^j$ (or that $M$ is scanning $\left(\alpha^j\right)_{i_j}$ on
the $j$-th tape); we also say that the $j$-th tape head is scanning the $i_j$-th tape square.

Let $I' = <q';\alpha^{1'}, ... , \alpha^{m'};i'_i, ... ,i'_m>$ be another instantaneous description
of $M$. We say that $I \rightarrow_M I'$ if $q \neq q_A, q \neq q_R$, and if there is
$s = < p;a_1, ... ,a_m;T_1, ... , T_m >$ in $\delta \left(q;\left(\alpha^1\right)_{i_1}, ... , \left(\alpha^m\right)_{i_m}\right)$ such that
$p = q'$, and, for each $j$, with $1 \leq j \leq m$:

\begin{enumerate}
\item $\left(\alpha^{j'}\right)_{i_j} = a_j$.
\item $\left(\alpha^{j'}\right)_k = \left(\alpha^j\right)_k$ for $1 \leq k \leq len\left(\alpha^j\right)$, if $k \neq i_j$.
\item $len\left(\alpha^{j'}\right) = len \left(\alpha^j\right)$ unless $i_j = len\left(\alpha^j\right) + 1$; in that case,\\ $len\left(\alpha^{j'}\right) = len\left(\alpha^j\right) + 1$.
\item If $T_j = L$, then $i_j \neq 1$.
\item If $T_j = R$, then $i' = i + 1$; if $T_j = L$, then $i'_j = i_j - 1$.
\end{enumerate}
We say that $M$ prints $a_j$ on the $j$-th tape. Note that $M$ cannot print a
blank (that is, $a_j \neq B$); so, we say that $\alpha^j$ is that portion of the $j$-th tape which
has been visited, or scanned. If $T_j = R\left(L\right)$, then we say that the $j-$th tape head
moves to the right (left). Assumption 4 corresponds to the intuitive notion of
each tape being one-way infinite to the right; thus, if $M$ "orders a tape head to
go off the left end of its tape," then $M$ halts. It is important to observe that
it is possible to have $I \rightarrow_M I_1$, and $I \rightarrow_M I_2$ with $I_1 \neq I_2$; hence the name
"nondeterministic."

We say $I \rightarrow_M^* J$ if there is a finite sequence $I_1, ... ,I_n$ such that
$I_1 = I, I_n = J$,and $I_i \rightarrow_M I_{i+1}$ for $1 \leq i < n$. Denote the empty string in $\Sigma^*$ by $\Lambda$. If $w \in \Sigma^*$, then let $\overline{w} = <q_0;w,\Lambda , ... ,\Lambda ;1, ... ,1>$ ($w$ is the
input). Call an instantaneous description $<q; \alpha_1, ... ,\alpha ; i_1, ... , i_m>$ \textit{accepting
}(\textit{rejecting}) if $q = q_A \left(q = q_R\right)$. We say that $M$ \textit{accepts} $w$ in $\Sigma^*$ if
$\overline{w} \rightarrow_M^* I$ for some accepting $I$. Denote by $A_M$, the set of all strings accepted
by $M$. We say that $M$ \textit{recognizes} $A_M$.

If $\overline{w} \rightarrow_M^* I$ for some accepting (rejecting) $I$, then we say that $M$, with
$w$ as input, eventually enters the accepting (rejecting) final state, and halts.

Intuitively speaking, there are three ways that a string $w$ in $\Sigma^*$ may be
not accepted by $M: M$, with $w$ as input, can eventually enter the rejecting final
state $q_R$; or $M$ can order a tape head to go off the left end of its tape; or $M$
can never halt.

Assume that $M$ is a multi-tape nondeterministic Turing machine, $w \in A_M$,
and $t$ is a positive integer. We say that \textit{$M$ accepts $w$ within $t$ steps} if, for
some $n \leq t$, 
\begin{align}
&there\ are\ instantaneous\ description\ I_1,\ ...\ ,I_{n+1} \nonumber \\
&such\ that\ I_1 = \overline{w}, I_{k+1}\ is\ accepting,\ and\ I_k\rightarrow_M I_{k+1} \label{eq:01} \\
&for\ 1 \leq k \leq n.  \nonumber
\end{align}
Let $s$ be a positive integer. Then \textit{$M$ accepts $w$ within space $s$} if for
some positive integer $n$, \eqref{eq:01} holds and, for each $I_k, 1 \leq k \leq n + 1$, if $I_k =$ $
<q;\alpha_1, ... ,\alpha^m;i_1, ... ,i_m>$, then $i_p \leq s$ for $1 \leq p \leq m$.

Let $T: N \rightarrow N$ and $S: N \rightarrow N$ be functions. We say that \textit{$M$ operates
in time $T$ (tape $S$)}, or \textit{$M$ recognizes $A_M$ in time $T$ (tape $S$)} if, fix each natural
number $l$ and each string $w$ in $A_m$, of length $l$, the machine $M$ accepts $w$
within $T(l)$ steps (space $S(1)$). We say that A is \textit{recognizable (non)deterministically in time $T$, or tape $S$}, if there is a multi-tape (non)deterministic Turing
machine $M$ that operates in time $T$, or tape $S$, such that $A = A_M$.

We will now define some well-known, important classes. Let $P\left(NP\right)$ be
the class of sets $A$ for which there is a positive integer $k$ such that $A$ is recognizable
(non)deterministically in time $l \mapsto l^k$. These are the \textit{(non)deterministic 
polynomial-time recognizable sets}.

Let $P_1\left(NP_1\right)$ be the class of sets $A$ for which there is a positive integer
$k$ such that $A$ is recognizable (non)deterministically in time $l \mapsto 2^{kl}$. These
are the \textit{(non)deterministic exponential-time recognizable sets}. If the positive
integer $n$ has length l in binary notation, then $2^{l-1} \leq n < 2l$. Therefore, a
set $A$ of positive integers is in $P_1\left(NP_1\right)$ iff there is a multi-tape (non)deterministic
Turing machine $M$, and a positive integer $k$ such that $A = A_M$, and $M$
accepts each $n$ in $A$ within $n^k$ steps. So in some sense, $P_1$ and $NP_1$ are
also classes of polynomial time recognizable sets.

We say that a set $A$ is recognizable in \textit{real time} if $A$ is recognizable in
time $I \mapsto l + 1$. We use $l + 1$ instead of $l$, so that the machine can tell when
it reaches the end of the input string.

We have defined Turing machines which recognize sets rather than compute
functions. It is clear how to modify our definitions to get the usual notion of a
function f computable by a deterministic one-tape Turing machine $M$; it is also
clear what we mean by $M$ \textit{computes the value of $f$ at $w$ within $t$ steps}. If
$f: A \rightarrow B$, where $A$ and $B$ are languages, and if $T: N \rightarrow N$, then we say that
\textit{$M$ computes $f$ in time $T$} if, for each natural number $1$ and each string $w$ in
$A$ of length $l$, the machine $M$ computes the value of $f$ at $w$ within $T\left(l\right)$
steps. 
\section{Notations of multivalued logic}
Let describe complex discrete logic units $z_n$, where $i, n \in \mathbb{Z}^+ \land i<n$ as 
\begin{equation}
z_n = e^{2 i \pi / n } 
\end{equation}
Let describe complex function $f^n\left(a,b\right), \forall a, b \in \left\lbrace 0, 1, 2, ... , n-1\right\rbrace $ as 
\begin{equation}
f^n\left(a,b\right) = {z_n}^{a \times b}
\end{equation}
where $\times$ denotes multiplication of two integers. Let describe complex function $g_k^n\left({z_n}^a\right), \forall a, \forall k \in \left\lbrace 0, 1, 2, ... , n-1\right\rbrace$ as 
\begin{equation}
g_k^n\left({z_n}^a\right) = {z_n}^{a + k}
\end{equation}
where $+$ denotes summation of two integers. 
\section{Lemmas of binary logic }
LEMMA 1. \textit{If $n = 2$, function $g_k^n\left({z_n}^a\right)$ is one argument binary logic generation function for binary set $ \left\lbrace {z_2}^0, {z_2}^1\right\rbrace $, where ${z_2}^0$ names \text{true} and ${z_2}^1$ names \text{false}}.\\
\textit{Proof}. The are $2^2$ different one argument logic functions:

\begin{align}
\varrho^{i_0}_{i_1}\left(a\right) &=
 \begin{array}{c|c}
  a & rez \\  
  \hline  
  \mbox{${z_2}^0$} & \mbox{$g^2_{i_0}\left({z_2}^a\right)$} \\
  \mbox{${z_2}^1$} & \mbox{$g^2_{i_1}\left({z_2}^a\right)$}
 \end{array}, \text{\ \ } \forall i_0, i_1 \in \left\lbrace 0, 1 \right\rbrace 
 \nonumber \\
\varrho^{0}_{0}\left(a\right) &=
 \begin{array}{c|c}
  a & rez \\  
  \hline  
  \mbox{${z_2}^0$} & \mbox{$g^2_0\left({z_2}^a\right)$} \\
  \mbox{${z_2}^1$} & \mbox{$g^2_0\left({z_2}^a\right)$}
 \end{array}
 =
  \begin{array}{c|c}
  a & rez \\  
  \hline  
  \mbox{${z_2}^0$} & \mbox{${z_2}^0$} \\
  \mbox{${z_2}^1$} & \mbox{${z_2}^1$}
 \end{array} \nonumber \\
\varrho^{0}_{1}\left(a\right) &=
\begin{array}{c|c}
  a & rez \\  
  \hline  
  \mbox{${z_2}^0$} & \mbox{$g^2_0\left({z_2}^a\right)$} \\
  \mbox{${z_2}^1$} & \mbox{$g^2_1\left({z_2}^a\right)$}
 \end{array}
 =
  \begin{array}{c|c}
  a & rez \\  
  \hline  
  \mbox{${z_2}^0$} & \mbox{${z_2}^0$} \\
  \mbox{${z_2}^1$} & \mbox{${z_2}^0$}
 \end{array} \label{eq:5} \\
\varrho^{1}_{0}\left(a\right) &= 
 \begin{array}{c|c}
  a & rez \\  
  \hline  
  \mbox{${z_2}^0$} & \mbox{$g^2_1\left({z_2}^a\right)$} \\
  \mbox{${z_2}^1$} & \mbox{$g^2_0\left({z_2}^a\right)$}
 \end{array}
 =
  \begin{array}{c|c}
  a & rez \\  
  \hline  
  \mbox{${z_2}^0$} & \mbox{${z_2}^1$} \\
  \mbox{${z_2}^1$} & \mbox{${z_2}^1$}
 \end{array} \nonumber \\
\varrho^{1}_{1}\left(a\right) &= 
 \begin{array}{c|c}
  a & rez \\  
  \hline  
  \mbox{${z_2}^0$} & \mbox{$g^2_1\left({z_2}^a\right)$} \\
  \mbox{${z_2}^1$} & \mbox{$g^2_1\left({z_2}^a\right)$}
 \end{array}
 =
  \begin{array}{c|c}
  a & rez \\  
  \hline  
  \mbox{${z_2}^0$} & \mbox{${z_2}^1$} \\
  \mbox{${z_2}^1$} & \mbox{${z_2}^0$}
 \end{array} \nonumber
 \end{align}
 
Direct calculations show, that $\varrho^{0}_{0}$ is self projection, $\varrho^{0}_{1}$ is antilogy, $\varrho^{1}_{0}$ is tautology, $\varrho^{1}_{1}$ is complementation.  $\bigcirc$ 
 
LEMMA 2. \textit{If $n = 2$, functions $f^n\left(a,b\right), g_k^n\left({z_n}^a\right)$ are two arguments binary logic generation functions for binary set $\left\lbrace {z_2}^0, {z_2}^1\right\rbrace $, where ${z_2}^0$ names \textit{true} and ${z_2}^1$ names \textit{false}}. 

\textit{Proof}. The are $2^{2^2}$ different two arguments logic functions:

 \begin{align}
\mu^{i_0,i_1}_{i_2,i_3}\left( a, b \right)
&=  
  \begin{array}{c|c c}
  a \backslash b &  \mbox{${z_2}^0$} & \mbox{${z_2}^1$}\\  
  \hline  
  \mbox{${z_2}^0$} & \mbox{$g_{i_0}\left( f\left( a, b \right) \right)$} & \mbox{$g_{i_1}\left( f\left( a, b \right) \right)$} \\
  \mbox{${z_2}^1$} & \mbox{$g_{i_2}\left( f\left( a, b \right) \right)$} & \mbox{$g_{i_3}\left( f\left( a, b \right) \right)$} 
 \end{array}, \text{\ \ } \forall i_0, i_1, i_2, i_3 \in \left\lbrace 0, 1 \right\rbrace  \nonumber \\
 \mu^{0,0}_{0,0}\left( a, b \right)
&=  
  \begin{array}{c|c c}
  a \backslash b &  \mbox{${z_2}^0$} & \mbox{${z_2}^1$}\\  
  \hline  
  \mbox{${z_2}^0$} & \mbox{${z_2}^0$} & \mbox{${z_2}^0$} \\
  \mbox{${z_2}^1$} & \mbox{${z_2}^0$} & \mbox{${z_2}^1$} 
 \end{array}, 
  \mu^{0,0}_{0,1}\left( a, b \right)
=  
  \begin{array}{c|c c}
  a \backslash b &  \mbox{${z_2}^0$} & \mbox{${z_2}^1$}\\  
  \hline  
  \mbox{${z_2}^0$} & \mbox{${z_2}^0$} & \mbox{${z_2}^0$} \\
  \mbox{${z_2}^1$} & \mbox{${z_2}^0$} & \mbox{${z_2}^0$} 
 \end{array} \nonumber \\
  \mu^{0,0}_{1,0}\left( a, b \right)
&=  
  \begin{array}{c|c c}
  a \backslash b &  \mbox{${z_2}^0$} & \mbox{${z_2}^1$}\\  
  \hline  
  \mbox{${z_2}^0$} & \mbox{${z_2}^0$} & \mbox{${z_2}^0$} \\
  \mbox{${z_2}^1$} & \mbox{${z_2}^1$} & \mbox{${z_2}^1$} 
 \end{array}, 
  \mu^{0,0}_{1,1}\left( a, b \right)
=  
  \begin{array}{c|c c}
  a \backslash b &  \mbox{${z_2}^0$} & \mbox{${z_2}^1$}\\  
  \hline  
  \mbox{${z_2}^0$} & \mbox{${z_2}^0$} & \mbox{${z_2}^0$} \\
  \mbox{${z_2}^1$} & \mbox{${z_2}^1$} & \mbox{${z_2}^0$} 
 \end{array} \nonumber \\
  \mu^{0,1}_{0,0}\left( a, b \right)
&=  
  \begin{array}{c|c c}
  a \backslash b &  \mbox{${z_2}^0$} & \mbox{${z_2}^1$}\\  
  \hline  
  \mbox{${z_2}^0$} & \mbox{${z_2}^0$} & \mbox{${z_2}^1$} \\
  \mbox{${z_2}^1$} & \mbox{${z_2}^0$} & \mbox{${z_2}^1$} 
 \end{array},  
  \mu^{0,1}_{0,1}\left( a, b \right)
=  
  \begin{array}{c|c c}
  a \backslash b &  \mbox{${z_2}^0$} & \mbox{${z_2}^1$}\\  
  \hline  
  \mbox{${z_2}^0$} & \mbox{${z_2}^0$} & \mbox{${z_2}^1$} \\
  \mbox{${z_2}^1$} & \mbox{${z_2}^0$} & \mbox{${z_2}^0$} 
 \end{array} \nonumber \\
  \mu^{0,1}_{1,0}\left( a, b \right)
&=  
  \begin{array}{c|c c}
  a \backslash b &  \mbox{${z_2}^0$} & \mbox{${z_2}^1$}\\  
  \hline  
  \mbox{${z_2}^0$} & \mbox{${z_2}^0$} & \mbox{${z_2}^1$} \\
  \mbox{${z_2}^1$} & \mbox{${z_2}^1$} & \mbox{${z_2}^1$} 
 \end{array},
  \mu^{0,1}_{1,1}\left( a, b \right)
=  
  \begin{array}{c|c c}
  a \backslash b &  \mbox{${z_2}^0$} & \mbox{${z_2}^1$}\\  
  \hline  
  \mbox{${z_2}^0$} & \mbox{${z_2}^0$} & \mbox{${z_2}^1$} \\
  \mbox{${z_2}^1$} & \mbox{${z_2}^1$} & \mbox{${z_2}^0$} 
 \end{array} \label{eq:6} \\
  \mu^{1,0}_{0,0}\left( a, b \right)
&=  
  \begin{array}{c|c c}
  a \backslash b &  \mbox{${z_2}^0$} & \mbox{${z_2}^1$}\\  
  \hline  
  \mbox{${z_2}^0$} & \mbox{${z_2}^1$} & \mbox{${z_2}^0$} \\
  \mbox{${z_2}^1$} & \mbox{${z_2}^0$} & \mbox{${z_2}^1$} 
 \end{array}, 
  \mu^{1,0}_{0,1}\left( a, b \right)
=  
  \begin{array}{c|c c}
  a \backslash b &  \mbox{${z_2}^0$} & \mbox{${z_2}^1$}\\  
  \hline  
  \mbox{${z_2}^0$} & \mbox{${z_2}^1$} & \mbox{${z_2}^0$} \\
  \mbox{${z_2}^1$} & \mbox{${z_2}^0$} & \mbox{${z_2}^0$} 
 \end{array} \nonumber \\
  \mu^{1,0}_{1,0}\left( a, b \right)
&=  
  \begin{array}{c|c c}
  a \backslash b &  \mbox{${z_2}^0$} & \mbox{${z_2}^1$}\\  
  \hline  
  \mbox{${z_2}^0$} & \mbox{${z_2}^1$} & \mbox{${z_2}^0$} \\
  \mbox{${z_2}^1$} & \mbox{${z_2}^1$} & \mbox{${z_2}^1$} 
 \end{array}, 
  \mu^{1,0}_{1,1}\left( a, b \right)
=  
  \begin{array}{c|c c}
  a \backslash b &  \mbox{${z_2}^0$} & \mbox{${z_2}^1$}\\  
  \hline  
  \mbox{${z_2}^0$} & \mbox{${z_2}^1$} & \mbox{${z_2}^0$} \\
  \mbox{${z_2}^1$} & \mbox{${z_2}^1$} & \mbox{${z_2}^0$} 
 \end{array} \nonumber \\
  \mu^{1,1}_{0,0}\left( a, b \right)
&=  
  \begin{array}{c|c c}
  a \backslash b &  \mbox{${z_2}^0$} & \mbox{${z_2}^1$}\\  
  \hline  
  \mbox{${z_2}^0$} & \mbox{${z_2}^1$} & \mbox{${z_2}^1$} \\
  \mbox{${z_2}^1$} & \mbox{${z_2}^0$} & \mbox{${z_2}^1$} 
 \end{array}, 
  \mu^{1,1}_{0,1}\left( a, b \right)
=  
  \begin{array}{c|c c}
  a \backslash b &  \mbox{${z_2}^0$} & \mbox{${z_2}^1$}\\  
  \hline  
  \mbox{${z_2}^0$} & \mbox{${z_2}^1$} & \mbox{${z_2}^1$} \\
  \mbox{${z_2}^1$} & \mbox{${z_2}^0$} & \mbox{${z_2}^0$} 
 \end{array} \nonumber \\
  \mu^{1,1}_{1,0}\left( a, b \right)
&=  
  \begin{array}{c|c c}
  a \backslash b &  \mbox{${z_2}^0$} & \mbox{${z_2}^1$}\\  
  \hline  
  \mbox{${z_2}^0$} & \mbox{${z_2}^1$} & \mbox{${z_2}^1$} \\
  \mbox{${z_2}^1$} & \mbox{${z_2}^1$} & \mbox{${z_2}^1$} 
 \end{array}, 
 \mu^{1,1}_{1,1}\left( a, b \right)
=  
  \begin{array}{c|c c}
  a \backslash b &  \mbox{${z_2}^0$} & \mbox{${z_2}^1$}\\  
  \hline  
  \mbox{${z_2}^0$} & \mbox{${z_2}^1$} & \mbox{${z_2}^1$} \\
  \mbox{${z_2}^1$} & \mbox{${z_2}^1$} & \mbox{${z_2}^0$} 
 \end{array} \nonumber 
 \end{align}
 
 Direct calculations show, that $\mu^{0,0}_{0,0}$ is nand, $\mu^{0,0}_{0,1}$ is antilogy, $\mu^{0,0}_{1,0}$ is left complementation, $\mu^{0,0}_{1,1}$ is if ... then, $\mu^{0,1}_{0,0}$ is right projection, $\mu^{0,1}_{0,1}$ is if, $\mu^{0,1}_{1,0}$ is neither ... nor, $\mu^{0,1}_{1,1}$ is if and only if (iff), $\mu^{1,0}_{0,0}$ is xor, $\mu^{1,0}_{0,1}$ is or, $\mu^{1,0}_{1,0}$ is not ... but, $\mu^{1,0}_{1,1}$ is right projection, $\mu^{1,1}_{0,0}$ is but not, $\mu^{1,1}_{0,1}$ is left projection, $\mu^{1,1}_{1,0}$ is tautology, $\mu^{1,1}_{1,1}$ is and \cite{bib:Knuth}. $\bigcirc$ 
 
\section{Lemmas of multivalued logic }

LEMMA 3. \textit{If $n > 2$, function $g_k^n\left({z_n}^a\right)$ is one argument multivalued logic generation function for multivalued set $ \left\lbrace {z_n}^0, {z_n}^1, {z_n}^2, .. , {z_n}^{n-1} \right\rbrace $}

\textit{Proof}. The are $n^n$ one argument logic functions:

\begin{align}
\varrho \begin{array}{l}
         \mbox{$i_0$} \\
         \mbox{$i_1$} \\
         \mbox{$i_2$} \\
         \dots \\
         \mbox{$i_{n-1}$} \end{array} \left(a\right) &=
 \begin{array}{l|l}
  a & rez \\  
  \hline  
  \mbox{${z_n}^0$} & \mbox{$g^n_{i_0}\left({z_n}^a\right)$} \\
  \mbox{${z_n}^1$} & \mbox{$g^n_{i_1}\left({z_n}^a\right)$} \\
  \mbox{${z_n}^2$} & \mbox{$g^n_{i_2}\left({z_n}^a\right)$} \\
  \dots & \dots \\
  \mbox{${z_n}^{n-1}$} & \mbox{$g^n_{i_{n-1}}\left({z_n}^a\right)$}
 \end{array}, \text{\ \ } \forall \begin{array}{l}
         \mbox{$i_0$} \\
         \mbox{$i_1$} \\
         \mbox{$i_2$} \\
         \dots \\
         \mbox{$i_{n-1}$} \end{array} \in \left\lbrace 0, 1, 2, ... , n-1 \right\rbrace  
\end{align}

All $\varrho$ function could be generated starting from index set $\left\{ i_0, i_1, i_2, ... , i_{n-1} \right\}$ $=$ $\left\lbrace 0, 0, 0, ... , 0 \right\rbrace$. For every two nearest $\varrho$ functions with index sets $\left\{ i_l, i_l, i_l,\right.$ $\left. ... , i_k, ... , i_l \right\}$ and $\left\{ i_l, i_l, i_l, ... , i_k+1, ... , i_l \right\}$ functions  $g^n_{i_l}\left({z_n}^a\right) = g^n_{i_l} \left({z_n}^a\right)$ and $g^n_{i_k}\left({z_n}^a\right) \neq g^n_{i_k+1} \left({z_n}^a\right)$. So all $n^n$ $\varrho$ functions with unique index set $\left\{ i_0, i_1, \right.$ $\left. i_2, ... , i_{n-1} \right\}$  are different. $\bigcirc$ 

LEMMA 4. \textit{If $n > 2$,  functions $f^n\left(a,b\right), g_k^n\left({z_n}^a\right)$ are two arguments multivalued logic generation functions for multivalued set $\left\lbrace {z_n}^0, {z_n}^1, {z_n}^2, ... , {z_n}^{n-1}\right\rbrace $}. 

\textit{Proof}. The are $n^{n^2}$ two arguments logic functions:

\begin{align}
\mu \begin{array}{lllll}
         \mbox{$i_{0,0}$} & \mbox{$i_{0,1}$} & \mbox{$i_{0,2}$} & \dots & \mbox{$i_{0, n-1}$} \\
         \mbox{$i_{1,0}$} & \mbox{$i_{1,1}$} & \mbox{$i_{1,2}$} & \dots & \mbox{$i_{1, n-1}$} \\
         \mbox{$i_{2,0}$} & \mbox{$i_{2,1}$} & \mbox{$i_{2,2}$} & \dots & \mbox{$i_{2, n-1}$} \\
         \dots & \dots & \dots & \dots & \dots \\
         \mbox{$i_{n-1, 0}$} & \mbox{$i_{n-1, 1}$} & \mbox{$i_{n-1, 2}$} & \dots & \mbox{$i_{n-1, n-1}$} \end{array}   &= \nonumber 
\end{align}
\begin{align} 
 \begin{array}{l|lllll}
  a \backslash b & \mbox{${z_n}^0$} & \mbox{${z_n}^1$} & \mbox{${z_n}^2$} & \dots & \mbox{${z_n}^{n-1}$} \\  
  \hline  
  \mbox{${z_n}^0$} & \mbox{$g^n_{i_{0,0}}\left({z_n}^{ab}\right)$} & \mbox{$g^n_{i_{0,1}}\left({z_n}^{ab}\right)$} & \mbox{$g^n_{i_{0,2}}\left({z_n}^{ab}\right)$} & \dots & \mbox{$g^n_{i_{0,n-1}}\left({z_n}^{ab}\right)$}\\
  \mbox{${z_n}^1$} & \mbox{$g^n_{i_{1,0}}\left({z_n}^{ab}\right)$} & \mbox{$g^n_{i_{1,1}}\left({z_n}^{ab}\right)$} & \mbox{$g^n_{i_{1,2}}\left({z_n}^{ab}\right)$} & \dots & \mbox{$g^n_{i_{1,n-1}}\left({z_n}^{ab}\right)$} \\
  \mbox{${z_n}^2$} & \mbox{$g^n_{i_{2,0}}\left({z_n}^{ab}\right)$} & \mbox{$g^n_{i_{2,1}}\left({z_n}^{ab}\right)$} & \mbox{$g^n_{i_{2,2}}\left({z_n}^{ab}\right)$} & \dots & \mbox{$g^n_{i_{2,n-1}}\left({z_n}^{ab}\right)$} \\
  \dots & \dots & \dots & \dots & \dots & \dots \\
  \mbox{${z_n}^{n-1}$} & \mbox{$g^n_{i_{n-1,0}}\left({z_n}^{ab}\right)$} & \mbox{$g^n_{i_{n-1,1}}\left({z_n}^{ab}\right)$} & \mbox{$g^n_{i_{n-1,2}}\left({z_n}^{ab}\right)$} & \dots & \mbox{$g^n_{i_{n-1,n-1}}\left({z_n}^{ab}\right)$}
 \end{array}, \nonumber
 \end{align}
 \begin{align}
  & \text{\ \ } \forall \begin{array}{lllll}
         \mbox{$i_{0,0}$} & \mbox{$i_{0,1}$} & \mbox{$i_{0,2}$} & \dots & \mbox{$i_{0, n-1}$} \\
         \mbox{$i_{1,0}$} & \mbox{$i_{1,1}$} & \mbox{$i_{1,2}$} & \dots & \mbox{$i_{1, n-1}$} \\
         \mbox{$i_{2,0}$} & \mbox{$i_{2,1}$} & \mbox{$i_{2,2}$} & \dots & \mbox{$i_{2, n-1}$} \\
         \dots & \dots & \dots & \dots & \dots \\
         \mbox{$i_{n-1, 0}$} & \mbox{$i_{n-1, 1}$} & \mbox{$i_{n-1, 2}$} & \dots & \mbox{$i_{n-1, n-1}$} \end{array} \in \left\lbrace 0, 1, 2, ... , n-1 \right\rbrace  
\end{align}
All $\mu$ function could be generated starting from index set\\ $\left\{ i_{0,0}, i_{0,1}, i_{0,2}, ... , i_{n-1,n-1} \right\} = \left\lbrace 0, 0, 0, ... , 0 \right\rbrace$. For every two nearest $\mu$ functions with index sets $\left\{ i_{l1,l2}, i_{l1,l2}, i_{l1,l2}, ... , i_{k1,k2}, ... , i_{l1,l2} \right\}$ and $\left\{ i_{l1,l2}, i_{l1,l2}, i_{l1,l2}, ... , i_{k1,k2}+1, ... , i_{l,l} \right\}$ functions $g^n_{i_{l1,l2}}\left({z_n}^{ab}\right) = g^n_{i_{l1,l2}} \left({z_n}^{ab}\right)$ and $g^n_{i_{k1,k2}}\left({z_n}^{ab}\right) \neq g^n_{i_{k1,k2}+1} \left({z_n}^{ab}\right)$. So all $n^{n^2}$ $\mu$ functions with unique index set $\left\{ i_{0,0}, i_{0,1}, i_{0,2}, ... , i_{n-1,n-1} \right\}$  are different. $\bigcirc$ 

LEMMA 5. \textit{Deterministic Touring machine counts symbols of string of length k in time $T = O\left( k^2 \right)$. Non-deterministic Touring machine  counts symbols of string of length k in time $T = O\left( k \right)$.} 

Touring machine string length counter working time could be expressed as 
\begin{align}
T(d) &= \underbrace{c_{iw} \left( \sum_{i=0}^{\frac{k}{d}-1} 2 d + \sum_{i=1}^{\frac{k}{d}-1} 2 i \right)}_{string\ walk}+\underbrace{c_{ow}\sum_{i=1}^{\log_d k} 2 i}_{counter\ walk} +\underbrace{c_{a} k}_{add\ one} \nonumber \\
  &= \left( 2 k + \left(\frac{k}{d}-1\right) \frac{k}{d} \right) c_{iw} + \log_d k \left( \log_d k +1 \right) c_{ow} +  k c_{a} \label{eq:time}
\end{align}
Working times of deterministic and non-deterministic Touring machines could be found by using of ~\eqref{eq:time} as follow 
\begin{align}
T\left(k\right) &= 2 k c_{iw} + k c_{a} = O\left( k \right) \label{eq:time1}\\
T\left(2\right) &= \left( 2 k + \left(\frac{k}{2}-1\right) \frac{k}{2} \right) c_{iw} + \log_2 k \left( \log_2 k +1 \right) c_{ow} + k c_{a} = O\left( k^2 \right) \label{eq:time2}
\end{align}
$\bigcirc$.
\begin{figure}
\centering
\includegraphics{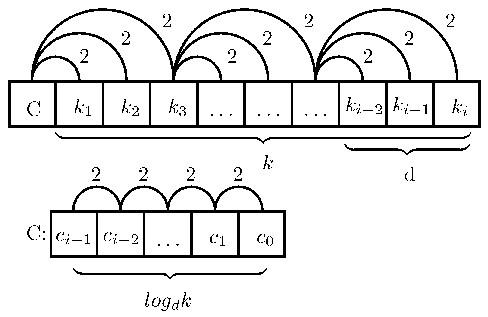}
\caption{Non-deterministic Touring counter machine}
\end{figure}

COROLLARY. \textit{The quickest counter is for $d=l$}

\textit{Proof}. If we choose $d=l$ our counter needs just one cell in tape for saving the count number in case $\log_l{l}=1$.
$\bigcirc$ 

%
%
\section{Proof of theorem for $k=6$}
TEOREM. \textit{There exist multivalued logic functions such that, for every countable function $T$
with $T\left(l\right) \geq l + 1$ for each l and for every language $A$ which is recognized
by a non-deterministic one-tape Turing machine in time $T$, the language $A$ is
recognized by a deterministic one-tape Turing machine in time $T^6$}.

\textit{Proof}. Every string $w$ of length $l$ in language $A$ could be expressed in non-deterministic one-tape Turing machine and in one-tape deterministic Turing machine so that 
\begin{align}
{z_{n_1}}^{\sum_{i=0}^{l-1}{b}^{i}\,{A_b}^{i}}&={z_{n_2}}^{\sum_{i=0}^{N_2-1}{2}^{i}\,{A_2}^{i}},\ for\ bases \label{eq:09}\\
 n_1&=b^l, b > 2, {A_b}^i < b, A_b^i \in {A}  \nonumber \\
 n_2&=2^{N_2}, {A_2}^i < 2, A_2^i \in {A} \nonumber
\end{align}
where $N_2$  length of string $w$ in $A_2$. If $b={2}^{{l}^{2}}$, so $N_2= l \log_2 \left(b\right)  = l^3$ and \eqref{eq:09} could be expressed as
\begin{align}
{z_{n_1}}^{\sum_{i=0}^{l-1}{b}^{i}\,{A_b}^{i}}&={z_{n_2}}^{\sum_{i=0}^{l^3-1}{2}^{i}\,{A_2}^{i}},\ for\ bases \label{eq:10}\\
 n_1&=2^{l^3}, {A_b}^i < b, {A_b}^i \in A \label{eq:11} \\
 n_2&=2^{l^3}, {A_2}^i < 2, {A_2}^i \in A \label{eq:12}
\end{align}
Let take the tape of non-deterministic Turing machine $M_{nd}$ of length $l+1+\log_{d=l}{l}$ and the tape of deterministic Turing machine $M_{d}$ of length $l^3+1+\log_{d=2}{l}$. Let write smallest integers in counter part of each Turing machine tape and all symbols of the string in language $A$ on the other part of tape. The  head of each Turing machine shifts from beginning of counter to new one symbol in the tape and back to the counter cell(s) at the beginning of the tape. Let choose for simplification reason one cell counter in $M_{nd}$ as described in previous subsection. Each time the head is at the beginning of counter it add 1. The non-deterministic Turing machine $M_{nd}$ needs $O(k)$ steps for that operation and deterministic Turing machine $M_{d}$ needs for that operation at least $O(k^2)$ steps for that operation. Let suppose that after reaching of last symbol of tape which is denoted as empty symbol Turing machine $M$ stops. Let use multivalued logic generation functions $g^{n_1}_k\left( {z_{n_1}}^{{A_b}^i} \right)$ where $b=2^{l^2}$ and $n_1$ described as \eqref{eq:11} for changing each symbol ${A_b}^i$ in language $A$ of non-deterministic Turing machine $M_{nd}$ and binary logic generation functions $g^{n_2}_k\left( {z_{n_2}}^{{A_2}^i} \right)$ where $n_2$ described as \eqref{eq:12} for changing each symbol ${A_2}^i$ in language $A$ of deterministic Turing machine $M_{d}$.  So, it follows from \eqref{eq:10}, that if non-deterministic Turing machine $M_{nd}$ recognize each symbol of language expressed as set $\left\{{A_b}^i \right\}, \forall i, 0 \le i \le l-1$ of length $l$ in time $T\left(l\right)\ge l + 1$, so deterministic Turing machine $M_{d}$ recognize the same symbols expressed in language $A$ as set $\left\{{A_2}^i \right\}, \forall i, 0 \le i \le l^3-1$ in time $T^6$ as follow from ~\eqref{eq:time1} and ~\eqref{eq:time2}. 
$\bigcirc$
\section*{Conclusion}
Proof of 2) proposition of Fagin's theorem 24 for $k=6$ implies  equivalence of propositions 2) and 1)  or $P = NP$.

\end{document}